\documentclass[prl,aps,amsfonts,amsmath,amssymb,nofootinbib,twocolumn,superscriptaddress]{revtex4}
\usepackage{mathtools}
\usepackage{upgreek}
\usepackage{lineno}
\usepackage{graphicx}
\usepackage{bm}
\usepackage{hyperref}
\usepackage{braket}
\newcommand{\RNum}[1]{\uppercase\expandafter{\romannumeral #1\relax}}
\usepackage{natbib}
\usepackage{float}
\usepackage{xcolor}

\restylefloat{table}

\begin{document}
\title{
Directionally asymmetric nonlinear optics in planar chiral MnTiO$_3$
}
\author{Xinshu Zhang}
\email{xszhang@physics.ucla.edu}
\affiliation{Department of Physics and Astronomy, University of California Los Angeles, Los Angeles, CA 90095, USA}

\author{Tyler Carbin}
\affiliation{Department of Physics and Astronomy, University of California Los Angeles, Los Angeles, CA 90095, USA}

\author{Kai Du}
\affiliation{Rutgers Center for Emergent Materials, Rutgers University, Piscataway, NJ, USA}

\author{Bingqing Li}
\affiliation{Rutgers Center for Emergent Materials, Rutgers University, Piscataway, NJ, USA}

\author{Kefeng Wang}
\affiliation{Rutgers Center for Emergent Materials, Rutgers University, Piscataway, NJ, USA}

\author{Casey Li}
\affiliation{Department of Physics and Astronomy, University of California Los Angeles, Los Angeles, CA 90095, USA}

\author{Tiema Qian}
\affiliation{Department of Physics and Astronomy, University of California Los Angeles, Los Angeles, CA 90095, USA}
\author{Ni Ni}
\affiliation{Department of Physics and Astronomy, University of California Los Angeles, Los Angeles, CA 90095, USA}

\author{Sang-Wook Cheong }
\affiliation{Rutgers Center for Emergent Materials, Rutgers University, Piscataway, NJ, USA}

\author{Anshul Kogar}
\email{anshulkogar@physics.ucla.edu}
\affiliation{Department of Physics and Astronomy, University of California Los Angeles, Los Angeles, CA 90095, USA}

\date{\today}

\maketitle

\textbf{
Planar chiral structures possess a two-dimensional handedness that is associated with broken mirror symmetry. Such motifs span vast length scales; examples include certain pinwheel molecules, nautilus shells, cyclone wind patterns and spiral galaxies. Although pervasive in nature, it has only recently been found that condensed matter systems can exhibit a form of planar chirality through toroidal arrangements of electric dipoles, known as ferro-rotational (FR) order. A key characteristic of such order is that enantiomorph conversion occurs when the solid is flipped by 180 degrees about an in-plane axis. 
Consequently, ferro-rotationally ordered materials may exhibit directionally asymmetric response functions, even while preserving inversion and time-reversal symmetry. Such an effect, however, has yet to be observed. Using second harmonic interferometry, we show here that when circularly polarized light is incident on MnTiO$_3$, the generated nonlinear signal exhibits directional asymmetry. Depending on whether the incident light is parallel or anti-parallel to the FR axis, we observe a different conversion efficiency of two right (left) circularly polarized photons into a frequency-doubled left (right) circularly polarized photon. Our work uncovers a fundamentally new optical effect in ordered solids and opens up the possibility for developing novel nonlinear and directionally asymmetric optical devices.}

Directional asymmetry (DA) refers to the breakdown of equivalence between the forward and backward direction of a physical process. Such asymmetry is present at the most fundamental levels of nature through the parity-violating weak interaction and also manifests itself at various other scales \cite{review1,review2,review3,review4,weak}. Examples include the Faraday effect, diode effect and magneto-chiral anisotropy effects, all of which give rise to the asymmetric propagation of electromagnetic signals \cite{faraday,MCA1,MCA2,SHGnonre1,SHGnonre2,SHGnonre3}.
In addition to their scientific importance, directional phenomena underpin devices such as semiconducting and superconducting diodes, optical isolators, circulators and acoustic gyrators \cite{diode1,diode2,diode3,acousticreview,isolator1,isolator2,isolator3,circulator1,circulator2}. In all of these cases, directional asymmetry arises in the presence of broken inversion ($\mathcal{I}$) and/or time-reversal ($\mathcal{T}$) symmetry, but DA does not require these symmetries to be broken. 

\begin{figure*}[t]
	\centering
	\includegraphics[width=1.8\columnwidth]{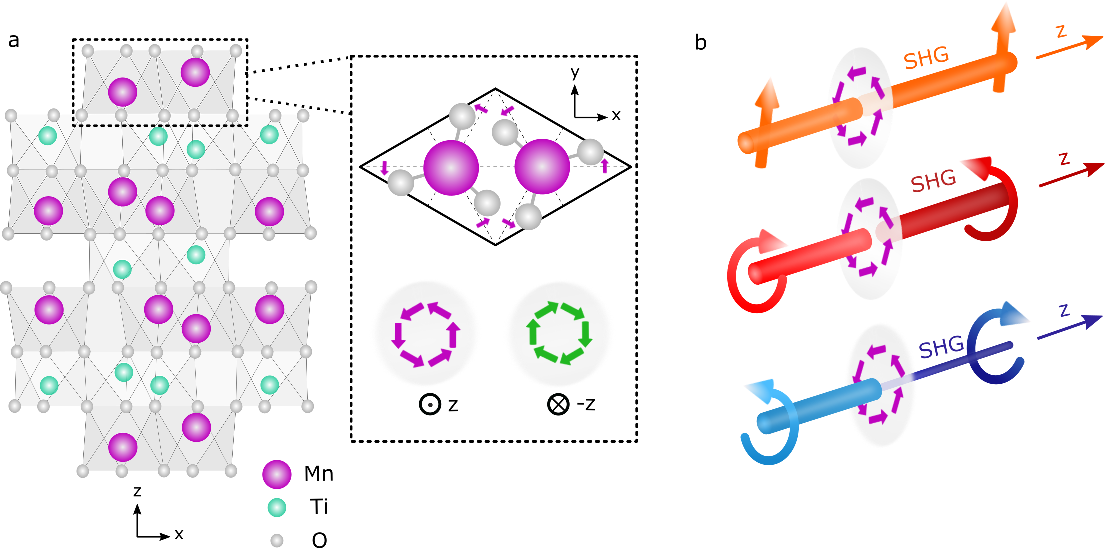}
	\caption{(a) The crystal structures of MnTiO$_3$ with a ferro-rotational space group (R$\bar{3}$). The buckled honeycomb layers consisting of Mn$^{2+}$ and Ti$^{4+}$ ions are stacked alternately along $z$ axis. Two oxygen cages enclosing two Mn$^{2+}$ ions are enlarged and the top view is shown. 
 Purple arrows denote the direction of rotational distortions of oxygen ions from the $\{110\}$ planes (dotted lines), breaking the mirror symmetry. The oxygen rotational displacements can be considered as planar chiral structures. The orientation of the FR order is counter-clockwise when viewing the structure along $z$ direction, but switches to the clockwise orientation if viewed from the $-z$ direction.
 (b) Schematic of second harmonic generation experiments, which include RA-SHG and C-SHG. For RA-SHG the incident and outgoing beam are linearly polarized along the same axis. For C-SHG, the incident and outgoing beam are circularly polarized with opposite helicities.}  
	\label{fig:1}
\end{figure*}

A fundamental question, therefore, is if a condensed matter system possessing appropriate order can support DA without breaking $\mathcal{I}$ or $\mathcal{T}$-symmetry. 
Because planar chiral structures exhibit a sense of twist that switches when viewed from the opposite direction 
(Fig.~\ref{fig:1}(a)), a material hosting such a structure should exhibit a directionally asymmetric effect. 
Pioneering works on engineered metamaterial structures possessing 
planar chirality have demonstrated the asymmetric transmission of circularly polarized microwaves \cite{meta1,meta2,meta3,meta4,meta5}. Since previous studies focused exclusively on two-dimentional metamaterials that were manually patterned with a specific twofold rotational symmetry axis, it remains unclear whether an analogous effect can be observed in a spontaneously ordered three-dimensional solid.
In this work, we show that materials hosting ferro-rotational (FR) order are ideal platforms for realizing DA without broken $\mathcal{I}$ or $\mathcal{T}$ symmetry.

To demonstrate the DA effect, we study the ferro-rotationally ordered compound MnTiO$_3$. At room temperature, the crystal structure of MnTiO$_3$ is represented by the centrosymmetric point group $\bar{3}$.  Mn$^{2+}$ and Ti$^{4+}$ ions in the buckled honeycomb layers are stacked alternately along the $z$ axis \cite{MTO1,MTO2,MTO3,MTO4,MTO5,MTO6,MTO7}.
Rotational distortions of oxygen cages around the Mn$^{2+}$ and
Ti$^{4+}$ ions break the $\{110\}$ mirror plane symmetry, which results in a toroidal arrangement of electric dipole moments, i.e. ferro-rotational order (Fig.~\ref{fig:1} (a)). In contrast to most ferro-rotational materials, which possess domains, MnTiO$_3$ naturally crystallizes in a single domain crystal which allows for unambiguous detection of directional asymmetry \cite{MTO6, FR1,FR2,FR3,FR4,FR5}. To reduce further experimental complexity, we restrict our investigation to temperatures above the Néel temperature ($T_N$=67~K) \cite{MTO2,MTO3,MTO4,MTO5,MTO6}. 

At present, most known ferro-rotationally ordered materials possess threefold rotational symmetry about the FR axis which forbids the asymmetric transmission of linear optical signals \cite{MTO6, FR1,FR2,FR3,FR4,FR5}. However, by extending the probe light into the nonlinear optical regime, we demonstrate that directional asymmetry of circularly polarized light can nevertheless be observed. We investigate MnTiO$_3$ with second harmonic generation (SHG), a process that converts two photons at frequency $\omega$ into one photon at frequency $2\omega$ ~\cite{SHGastool,Boyd,Powell}. MnTiO$_3$ possesses inversion symmetry, which forbids the lowest order electric dipole contribution to SHG. The sub-leading contributions, magnetic dipole and electric quadrupole SHG, are both allowed and cannot be distinguished in our experiment; for simplicity, we refer only to magnetic dipole SHG in the remainder of the text. 

To boost this usually weak signal, we tune the wavelength of the fundamental light between 960 and 1030~nm so that the SHG energy resonates with $d$-$d$ transitions in the energy range spanning 2.4-2.6~eV \cite{SHGastool,MTO5}. In the ferro-rotational state, two independent susceptibility tensor elements are permitted, $\chi_{yyy}=-\chi_{yxx}=-\chi_{xyx}=-\chi_{xxy}$ and $\chi_{xxx}=-\chi_{xyy}=-\chi_{yxy}=-\chi_{yyx}$, but only the $\chi_{yyy}$-based terms are proportional to the ferro-rotational order parameter such that $\chi_{yyy}(\textbf{L}) = -\chi_{yyy}(-\textbf{L})$, where $\textbf{L}$ is the order parameter~\cite{MTO5,MTO6,FR1,Birss} (Supplementary Note~\RNum{1}). The presence of both terms, with only $\chi_{yyy}$ dependent on the sign of $\textbf{L}$, ensures sensitivity to the orientation of the FR order through interference. The relative magnitudes and phases of $\chi_{xxx}$ and $\chi_{yyy}$ can be extracted by rotational anisotropy second harmonic generation (RA-SHG). In our experiment, the incident fundamental and detected second harmonic light are linearly polarized along the same direction. Once the relevant parameters have been extracted, we then perform circularly polarized second harmonic generation (C-SHG) with light propagating along the toroidal axis to illustrate intensity contrast between left and right circularly polarized light as well as the directional asymmetry effect. 


\begin{figure*}[t]
	\centering
	\includegraphics[width=1.98\columnwidth]{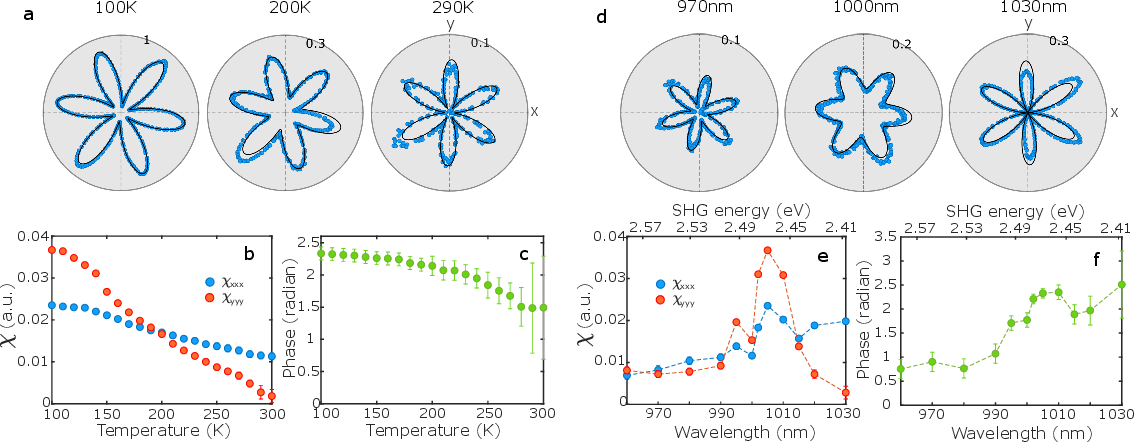}
	\caption{(a) RA-SHG patterns with incident light wavelength 1005 nm at 100, 200 and 290 K . The intensities of all datasets are normalized to a value of 1 corresponding to 100 K. (b)  $\chi_{xxx}$ (blue) and $\chi_{yyy}$ (red) as a function of temperature from 100~K to 300 K. (c) The relative phase $\gamma$ between $\chi_{xxx}$ and $\chi_{yyy}$ as a function of temperature. (d) RA-SHG patterns for three selected incident wavelengths 970, 1000 and 1030 nm at 100 K. The intensities of all datasets are normalized to a value of 1 corresponding to 1005 nm at 100 K. (e)  $\chi_{xxx}$ and $\chi_{yyy}$ as a function of wavelength from 960 to 1030 nm. (f) The relative phase $\gamma$ between $\chi_{xxx}$ and $\chi_{yyy}$ as a function of wavelength. The error bars represent the standard deviation from the fit.  }  
	\label{fig:2}
\end{figure*}

To quantify the tensor elements, we first present RA-SHG measurements as a function of temperature at a wavelength of 1005~nm (Fig.~\ref{fig:2}(a)). (RA-SHG patterns at a greater number of temperature points are shown in Supplementary Note~\RNum{3}). The RA pattern at all temperatures can be fit with a simple expression (Supplementary Note~\RNum{1}):
\begin{equation}
    I(2\omega) \propto |\chi_{xxx}\textrm{sin}(3\theta) + e^{i\gamma} \chi_{yyy}\textrm{cos}(3\theta)|^2 \\
    \label{eq:RA-SHG}
\end{equation}
where $\chi_{xxx}$ and $\chi_{yyy}$ are the real-valued magnetic dipole tensor elements, $\gamma$ denotes the relative phase between the two tensor elements and 
$\theta$ represents the polarization angle of the incident and detected light with
respect to the $x$ axis in MnTiO$_3$ (Fig.~\ref{fig:1} (a)). At room temperature, the magnitude of $\chi_{yyy}$ is negligible compared to $\chi_{xxx}$ and the RA-SHG pattern exhibits almost perfect nodes 
(Supplementary Note~\RNum{2}). As the temperature is lowered, the magnitude of $\chi_{yyy}$ increases. Consequently, the RA pattern rotates counterclockwise away from the crystal axes and the nodes are lifted as shown in Fig.~\ref{fig:2} (a). The lifted nodes shrink again at low temperatures because $\chi_{yyy}$ becomes significantly larger than $\chi_{xxx}$. By fitting the RA-SHG patterns, we can extract the magnitudes of $\chi_{xxx}$ and $\chi_{yyy}$ in addition to the relative phase as a function of temperature (Fig.~\ref{fig:2} (b)-(c)). 

The $\chi_{xxx}$/$\chi_{yyy}$ ratio and the relative phase $\gamma$ depend not only  on the temperature, but also sensitively  on the wavelength of the incident light. By scanning the incident light over a wavelength of 960-1030~nm, which corresponds to a $^6A_{1g}$$\rightarrow$$^4T_{2g}$ crystal field transition, $\gamma$ changes from roughly $\pi/4$ to 3$\pi$/4, while $\chi_{xxx}$ and $\chi_{yyy}$ also change significantly (Fig.~\ref{fig:2} (e)-(f)). Consequently, the RA-SHG patterns also adjust accordingly as shown in Fig.~\ref{fig:2} (d) (more RA-SHG patterns at various wavelengths are shown in Supplementary Note~\RNum{4}). Over our experimental range of wavelengths and temperatures, the relative phase does not switch sign, which, as we show below, would correspondingly invert the contrast between left and right circularly polarized light. Nevertheless, it should be noted that nothing forbids such a phase change, and over a different energy range, this contrast may be reversed.

\begin{figure}[b]
	\includegraphics[width=0.96\columnwidth]{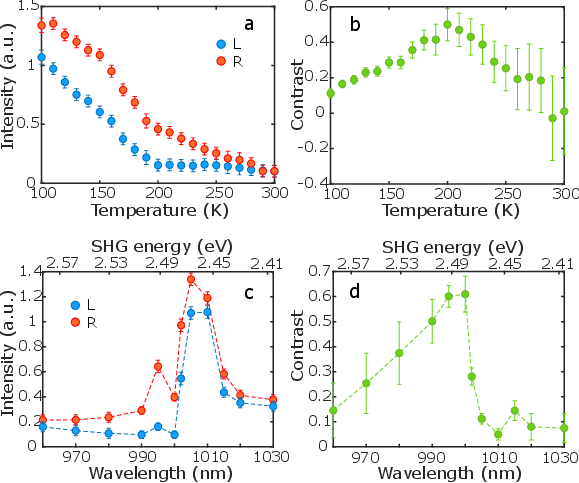}
	\caption{  (a) The intensity of C-SHG as a function of temperature for incident light with left (L) and right (R) circular polarization at an incident wavelength of 1005~nm. (b) The temperature dependent C-SHG contrast defined as (R-L)/(R+L), highlighting the different efficiencies for left and right circularly polarized incident light. (c)  The intensity of C-SHG as a function of wavelength for incident light with left (L) and right (R) circular polarization at $T=100$~K.  (d) The wavelength dependent C-SHG contrast. The error bars represent the standard deviation from two independent measurements.  }  
	\label{fig:3}
\end{figure} 

\begin{figure*}[t]
	\centering
	\includegraphics[width=1.96\columnwidth]{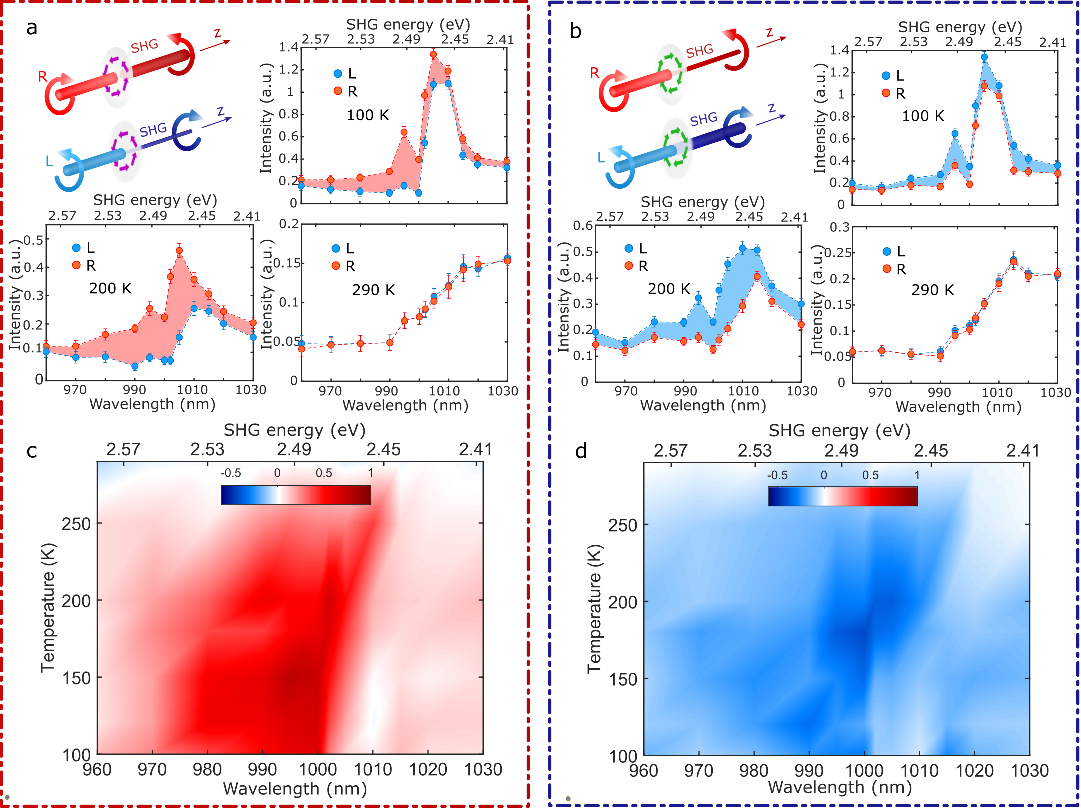}
\caption{(a)-(b) Wavelength dependent C-SHG  for two planar chiral structures at 100~K, 200~K and 290~K. The error bars represent the standard deviation from two independent measurements. (c)-(d) Color plots depicting the normalized SHG intensity contrast between left and right circularly polarized incident light as a function of temperature and wavelength for two planar chiral structures.  }  
	\label{fig:4}
\end{figure*} 

Because of the aforementioned interference, left and right circularly polarized light generate the second harmonic at different efficiencies. The efficiencies sensitively depend on the temperature and wavelength as shown in Figs.~\ref{fig:3} (a) and (c). SHG intensity contrast between left and right circularly polarized incident light is shown in Figs.~\ref{fig:3} (b) and (d). When the incident wavelength is 1005~nm, the contrast is not visible at room temperature within our experimental signal-to-noise ratio because the magnitude of $\chi_{yyy}$ is small compared to $\chi_{xxx}$. But as $\chi_{yyy}$ becomes of equal magnitude to $\chi_{xxx}$ at roughly 200~K (Fig.~\ref{fig:2} (b)), the contrast is maximal~(Fig.~\ref{fig:3} (b)). Understanding this disparity requires converting Eq.~\ref{eq:RA-SHG} to the circular basis: 
\begin{equation}
\begin{split}
 &I_{L}(2\omega) \propto |i\chi_{xxx} + e^{i\gamma} \chi_{yyy}|^2 I_{R}^2(\omega)\\
& I_{R}(2\omega) \propto |i\chi_{xxx} - e^{i\gamma} \chi_{yyy}|^2 I_{L}^2(\omega).
\label{C-SHG}
\end{split}
\end{equation}
From these expressions, it is clear that the contrast between the right and left circularly polarized light is observable only when the matrix elements are of comparable magnitude and when $\gamma \neq m\pi$, where $m$ is an integer. In correspondence with the RA patterns, the circularly polarized contrast is most pronounced when the node is maximally lifted. At 100~K, when the wavelength is tuned between 995-1000~nm, $\gamma \approx \pi/2$ and $|\chi_{xxx}| \approx |\chi_{yyy}|$, which means that primarily left circularly polarized second harmonic light is generated, while right circularly polarized light is hardly produced. The contrast is thus maximal at this point, as shown in Figs.~\ref{fig:3}~(d). It is important to note that while the contrast between left and right circularly polarized light is necessary to observe the directional asymmetry effect, it is not a sufficient condition (e.g. circularly polarized SHG contrast is observed in Cr$_2$O$_3$ without directional asymmetry) \cite{Cr2O3,Cr2O3Fiebig,topography}.  

\begin{figure*}[t]
	\centering
	\includegraphics[width=1.95\columnwidth]{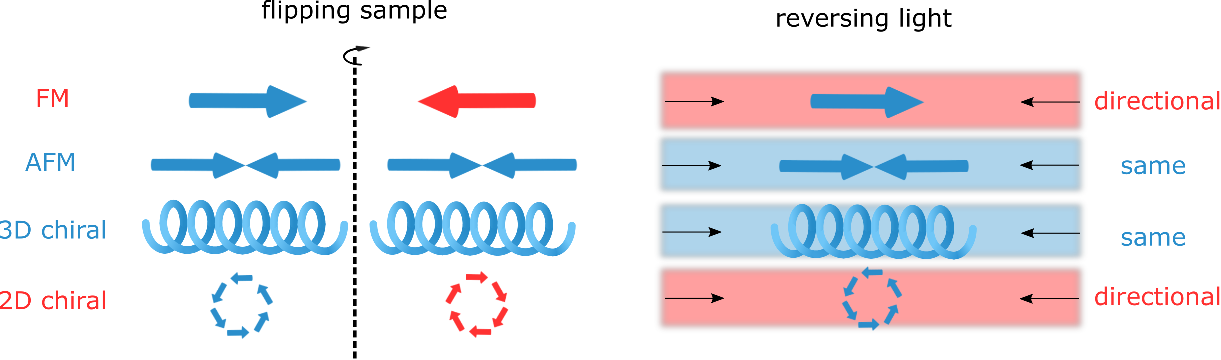}
\caption{Four representative systems are chosen to illustrate the conditions required for directional asymmetry, including a ferromagnet (FM), an anti-ferromagnet (AFM), a three dimensional (3D) chiral system and a planar (2D) chiral system. Due to the lack of in-plane $C_2$ symmetry, flipping the FM and the planar chiral system results in distinct properties (e.g., the spin direction reverses in the FM while a counter-clockwise rotation switches to a clockwise rotation for a planar chiral system). Consequently, circularly polarized light propagating in opposite directions interact differently with these material systems. In contrast, in AFMs and 3D chiral systems, which preserve $C_2$ symmetry and remain unchanged upon flipping, the same electromagnetic response is observed regardless of the direction of propagation.}  
	\label{fig:5}
\end{figure*} 

The presence of directional asymmetry in the ferro-rotational state of MnTiO$_3$ is illustrated in Fig.~\ref{fig:4}. In Fig.~\ref{fig:4} (a), wavelength-dependent C-SHG is shown for one planar chiral structure at 100~K, 200~K and 290~K, while Fig.~\ref{fig:4} (b) shows the results for the other enantiomorph by flipping the sample 180 degrees about an in-plane axis. Flipping the sample in this way is equivalent to reversing the positions of the source and detector and therefore comprises a test of directional asymmetry. Prior to flipping the sample, we observe a higher second harmonic conversion efficiency for right circularly polarized light, whereas afterwards, the same is true of left circularly polarized light. As expected from the tensor elements and relative phase extracted from the RA-SHG, the DA contrast is largest at around 200~K, while asymmetry can still be observed at 100~K. At 290~K, however, because $\chi_{yyy}$ is negligibly small compared to $\chi_{xxx}$, we do not observe contrast between left and right circularly polarized light; subsequently, the directional asymmetry effect disappears. This observation illustrates that the directional asymmetry effect arises due to interference between tensor elements that are proportional to the order parameter and elements that are not. Only the $\chi_{yyy}$-based elements change sign upon rotating the sample, while the $\chi_{xxx}$ terms provide an unchanged reference. Figs.~\ref{fig:4} (c) and (d) summarize the directional effect as a function of wavelength and temperature, revealing a clear directional asymmetry  across nearly the entire range of the relevant phase space.

To understand the general requirements for observing the DA effect with circularly polarized light, one must consider the roles played by symmetry and angular momentum transfer. Crucially, twofold rotational symmetry about any axis perpendicular to the FR axis must be broken. Colloquially stated, it is imperative that flipping the sample over will not leave the crystal invariant. As we show in Fig.~\ref{fig:5}, this constraint is also obeyed by Faraday media, which also exhibit directional asymmetry in the propagation speed of circularly polarized light. On the other hand, simple collinear anti-ferromagnetic systems (like Cr$_2$O$_3$) and three dimensional chiral systems, which possess a twofold in-plane rotational symmetry axis, exhibit the same response for counter-propagating light waves \cite{Cr2O3,Cr2O3Fiebig,topography,activity}. However, the absence of this rotational axis is not the only relevant broken symmetry. Specifically, all mirror planes that contain the FR axis must also be broken. The presence of such mirror symmetries would prevent a preference for one circular polarization over the other, which is essential for the observed effect. Notably, neither inversion symmetry nor time-reversal symmetry need to be absent; the relevant twofold rotation and mirror operations are the primary symmetries to be considered for observing the DA effect presented here.

Whether one observes this DA effect within the realm of linear or nonlinear optics depends on considerations related to angular momentum conservation. An important aspect of Eq.~\ref{C-SHG} is that two left (right) circularly polarized incident photons are converted into one right (left) circularly polarized second harmonic photon. Each circularly polarized photon carries  an angular momentum of $\pm\hbar$; an incident photon will transfer this amount to a solid, while an outgoing photon will carry it away. Thus, the conversion of photons from right (left) circularly polarized light to left (right) circularly polarized light entails elastically transferring angular momentum to the solid. Such a process is analogous to that of Bragg diffraction from a crystal, where linear momentum (but not energy) is transferred. The amount of angular momentum that can be transferred to the solid with circularly polarized light depends on the rotational symmetry of the crystal about the propagation axis \cite{angular,CrI3}. As established  in Ref.~\cite{angular}, integer multiples of $ n\hbar$ units of angular momentum, where  $n$ is an integer, can be transferred to a crystal when the light propagates along an $n$-fold rotational symmetry axis. In the case of MnTiO$_3$, where light propagates along a threefold symmetric rotation axis, the process in which two right (left) circularly polarized photons incident on the crystal produce a single outgoing left (right) circularly polarized photon at twice the frequency (which transfers $\pm3\hbar$ units of angular momentum to the solid) is therefore allowed. Consequently, the directional asymmetry effect in MnTiO$_3$ 
cannot be observed in a linear optical experiment, which would require the FR axis to be a twofold rotational symmetry axis. A second order nonlinear optical process is thus the key to the helicity conversion of circularly polarized photons and underlies the directional asymmetry effect in MnTiO$_3$.

Our observations demonstrate that directional asymmetry of circularly polarized light can be realized in a system without broken $\mathcal{I}$ or $\mathcal{T}$-symmetry in ferro-rotational MnTiO$_3$. We predict that solids sharing a twofold rotational symmetry axis and a toroidal axis due to ferro-rotational order should exhibit a directionally asymmetric conversion efficiency of circularly polarized light without invoking nonlinear optics.  
Furthermore, engineered two-dimensional materials with Moire superstructures that produce planar chiral motifs also offer promising avenues through which to  investigate directionally asymmetric propagation of circularly polarized light \cite{TMD1,TMD2,TaS2}. Our findings pave the way for the search of directional asymmetry in other unconventional structures and expand 
 the material database for the development of novel linear and nonlinear optical devices.

\nolinenumbers
\footnotesize
\vspace{-0.75em}
 \section{Acknowledgements:}
We thank Xianghan Xu and Boxuan Zhou for helpful conversations related to this work. We thank Jonathan Loera and J Green for the help in X-ray diffraction measurements.
The SHG experiments at UCLA were supported by the U.S. Department of Energy (DOE), Office of Science, Office of Basic Energy Sciences under Award No. DE-SC0023017. N.N. and T.Q. were supported by the U.S. Department
of Energy (DOE), Office of Science, Office of Basic Energy Sciences under Award Number DE-SC0021117. The work at Rutgers was supported by W. M. Keck Foundation. 

\section{Author contributions:  }
X.Z. performed the RA-SHG and C-SHG experiments with the help from T.C. and C.L.. X.Z. analysed the data. The SHG project was performed under the supervision of A.K..  K.D., B.L. and K.W. grew the single crystals under the supervision of S.-W.C.. T.Q. and N.N. prepared the crystal for measurements and characterized the crystal with X-ray diffraction. The manuscript was written by X.Z. and A.K. with input from all authors.
\vspace{-0.75em}

\section{Competing interests:  }

The authors declare no competing interests.

\section{Methods }
\vspace{-1em}
\subsection{Sample synthesis}
MnO (Thermo Scientific, 99.99$\%$) and TiO$_2$ (Alfa Aesar, 99.8$\%$) powder were mixed at the stoichiometric ratio and calcined in air from 1000$^{\circ}$C to 1400$^{\circ}$C with intermediate grindings at each 100$^{\circ}$C interval. The powder was then made into feed-rods that has been hydrostatically pressed at 800 bar and sintered at 1400$^{\circ}$C for 20 hours. Single crystals of MnTiO$_3$ were grown in the laser floating-zone furnace (Crystal System Corporation, Japan) where the feed-rods were grown at 3 mm/h in Argon with ambient pressure and both feed and seed rods were counter-rotated at rates around 18 rpm.

\subsection{Experimental details}
The regeneratively amplified laser used in our experiment is based on a Yb:KGW gain medium that outputs a power of 10~W. The laser pulses have a Gaussian-like profile with an approximately 180~fs pulse duration and a 1030~nm central wavelength. In our experiment, we used a laser pulse repetition rate of 20~kHz.  The laser pulse was generated from an optical parametric amplifier with tunable wavelength, which we use for the second harmonic spectroscopy between 960-1030~nm. The laser pulse was focused normally on the sample with a 100~$\mu$m spot size, and the probe fluence was approximately 15~$\mathrm{mJ}/\mathrm{cm}^{2}$. Detection of the second harmonic light was conducted with a commercial photo-multiplier tube. The sample was cooled  with a standard optical cryostat with fused silica windows to prevent distortions to the light polarization.

 \section{Data availability}
 \vspace{-1em}
 The data that supports the findings of this study are present in the paper and/or in the supplementary information, and are deposited in the Zenodo repository.  Additional data related to the paper is available from the corresponding authors upon reasonable request.

\normalsize

\end{document}



\newcommand{\pd}{\partial}
\newcommand{\beq}{\begin{equation}}
\newcommand{\eeq}{\end{equation}}
\newcommand{\bseq}{\begin{subequations}}
\newcommand{\eseq}{\end{subequations}}
\newcommand{\bpmat}{\begin{pmatrix}}
\newcommand{\epmat}{\end{pmatrix}}
\title{Supplementary Information for Directionally asymmetric nonlinear optics in planar chiral MnTiO$_3$}
\author{Xinshu Zhang}
\email{xszhang@physics.ucla.edu}
\affiliation{Department of Physics and Astronomy, University of California Los Angeles, Los Angeles, CA 90095, USA}

\author{Tyler Carbin}
\affiliation{Department of Physics and Astronomy, University of California Los Angeles, Los Angeles, CA 90095, USA}

\author{Kai Du}
\affiliation{Rutgers Center for Emergent Materials, Rutgers University, Piscataway, NJ, USA}

\author{Bingqing Li}
\affiliation{Rutgers Center for Emergent Materials, Rutgers University, Piscataway, NJ, USA}

\author{Kefeng Wang}
\affiliation{Rutgers Center for Emergent Materials, Rutgers University, Piscataway, NJ, USA}

\author{Casey Li}
\affiliation{Department of Physics and Astronomy, University of California Los Angeles, Los Angeles, CA 90095, USA}

\author{Tiema Qian}
\affiliation{Department of Physics and Astronomy, University of California Los Angeles, Los Angeles, CA 90095, USA}
\author{Ni Ni}
\affiliation{Department of Physics and Astronomy, University of California Los Angeles, Los Angeles, CA 90095, USA}

\author{Sang-Wook Cheong }
\affiliation{Rutgers Center for Emergent Materials, Rutgers University, Piscataway, NJ, USA}

\author{Anshul Kogar}
\email{anshulkogar@physics.ucla.edu}
\affiliation{Department of Physics and Astronomy, University of California Los Angeles, Los Angeles, CA 90095, USA}

\date{\today}

\maketitle
\section{Analysis of SHG }
\subsection{Rotational anisotropy SHG}
Radiation can come from various sources, including electric dipoles, magnetic dipoles, electric quadrupoles, and higher order radiation sources. In MnTiO$_3$ above anti-ferromagnetic transition temperature $T_N=67 $K, the electric dipole contribution to SHG is forbidden due to inversion symmetry, while the magnetic dipole and electric quadrupole contributions are allowed. These two contributions give rise to the same expressions in our experimental geometry and therefore yield no additional information pertaining to the symmetry of the emitted radiation. Therefore, for simplicity, we neglect the electric quadrupole contribution and consider only the magnetic dipole SHG. The source term entering Maxwell's equations is then given by 
$\boldsymbol{S}=\nabla \times \frac{\partial  \boldsymbol{M}}{\partial t}$. 
The magnetic contribution at the second harmonic frequency has the form of $\boldsymbol{M}(2\omega)=\boldsymbol{\chi}\boldsymbol{E}(\omega)\boldsymbol{E}(\omega)$, where $\boldsymbol{\chi}$ is the second order nonlinear magnetic susceptibility tensor, and $\boldsymbol{E}$ is the incident electric field at the fundamental frequency. Above $T_N$, MnTiO$_3$ has point group $\bar{3}$. For light propagating along the $z$-direction (optical axis of the crystal), the magnetic susceptibility tensors have only two independent tensor elements with the relation $\chi_{yyy}$=$-\chi_{yxx}$=$-\chi_{xyx}$=$-\chi_{xxy}$ and $\chi_{xxx}$=$-\chi_{xyy}$=$-\chi_{yxy}$=$-\chi_{yyx}$.  Therefore, the source term is given by:
\begin{equation}
\boldsymbol{S} \propto \nabla \times \frac{\partial \boldsymbol{M}}{\partial t}\propto 
\begin{pmatrix}
\chi_{yyy}(E_{x}^{2}-E_{y}^{2})+2\chi_{xxx}E_{x}E_{y}  \\
\chi_{xxx}(E_{x}^{2}-E_{y}^{2})-2\chi_{yyy}E_{x}E_{y}  \\
0 
\end{pmatrix}.
  \label{eq:S1}
\end{equation}
The second harmonic intensity measured by the detector in our experiment is $I \propto |\textbf{e}(2\omega)\cdot \boldsymbol{S} |^2$, where  $\textbf{e}(2\omega)$ is the unit vector of the output polarizer that selects second harmonic light of a particular polarization.   By taking the polarization of both the incident fundamental light and the outgoing SHG light into account, we obtain \begin{equation}
    I(2\omega) \propto |\chi_{xxx}\textrm{sin}(3\theta) +  \chi_{yyy}\textrm{cos}(3\theta)|^2, \\
    \label{eq:S2}
\end{equation} where $\theta$ is taken to be with respect to the $x$-axis.   $\chi_{yyy}$ and $\chi_{xxx}$ are complex quantities, but only the relative phase between them is important. We can rewrite the above expression as \begin{equation}
    I(2\omega) \propto \mid|\chi_{xxx}| \textrm{sin}(3\theta) +  e^{i\gamma}|\chi_{yyy}| \textrm{cos}(3\theta)\mid^2, \\
    \label{eq:S3}
\end{equation}  The magnitude $|\chi_{yyy}|$, $|\chi_{xxx}|$ and relative phase $\gamma$ can be obtained by fitting to the rotational anisotropy second harmonic generation (RA-SHG) patterns.

\subsection{Circular SHG}

In this section, we demonstrate that illuminating MnTiO$_3$ with left circularly polarized light results in the emission of right circularly polarized light at twice the frequency, and vice versa. This is a consequence of angular
momentum conversion and the threefold rotational symmetry of the crystal about the $z$-axis of the crystal. 
We can convert the source term from linear basis in~\ref{eq:S1}  into circular basis with $\boldsymbol{E}=E_{R}\textbf{e}_{R}+E_{L}\textbf{e}_{L}+E_{z}\textbf{e}_{z}$, where $
 \textbf{e}_{R}=-1/\sqrt{2}(\textbf{e}_{x}+i \textbf{e}_{y})$ and $\textbf{e}_{L}=1/\sqrt{2}(\textbf{e}_{x}-i \textbf{e}_{y})$. Then one gets 
 
 \begin{equation}
\boldsymbol{S}=\begin{pmatrix}
S_{R}\\
S_{L}\\
S_{z}
\end{pmatrix}
\propto  
\begin{pmatrix}
(-\chi_{yyy}+i\chi_{xxx}) E_{L}^{2} \\
(\chi_{yyy}+i\chi_{xxx}) E_{R}^{2}  \\
0 
\end{pmatrix}
 \label{eq:S4}
\end{equation}
 $ I_{R/L} \propto  |\boldsymbol{S}|^{2}$. Hence, the circular SHG intensity is 
 \begin{equation}
\begin{split}
 &I_{L}(2\omega) \propto |\chi_{yyy} + i\chi_{xxx}|^2 I_{R}^2(\omega)\\
& I_{R}(2\omega) \propto |-\chi_{yyy} + i\chi_{xxx}|^2 I_{L}^2(\omega).
\label{eq:S5}
\end{split}
\end{equation}
Alternatively, we can express these expressions in terms of the magnitude and phase:
 \begin{equation}
 \begin{split}
    &I_{L} \propto  \left(|\chi_{xxx}|^2+|\chi_{yyy}|^2\right)+ 2\left(|\chi_{xxx}| |\chi_{yyy}|\right)\textrm{sin}(\gamma )\\
    &I_{R} \propto  \left(|\chi_{xxx}|^2+|\chi_{yyy}|^2\right)- 2\left(|\chi_{xxx}| |\chi_{yyy}|\right)\textrm{sin}(\gamma ).
    \label{eq:S6}
    \end{split}
   \end{equation} The first term is quadratic in $\chi_{xxx}$ and $\chi_{yyy}$,  and is always positive, while the second term is the so-called interference term and can be either positive or negative. The sign of the interference term can be reversed by either changing the handedness of the circular light or by flipping the sample 180 degrees. Additionally, the interference is also affected by the wavelength of the light. The greatest contrast in the second harmonic response to the incident left and right circularly polarized light occurs when the  $\chi_{xxx}$ and $\chi_{yyy}$ have equal magnitudes and possess a 90-degree phase difference.

\section{Room temperature RA-SHG }
\begin{figure}[hbt!]
	\includegraphics[width=0.9\columnwidth]{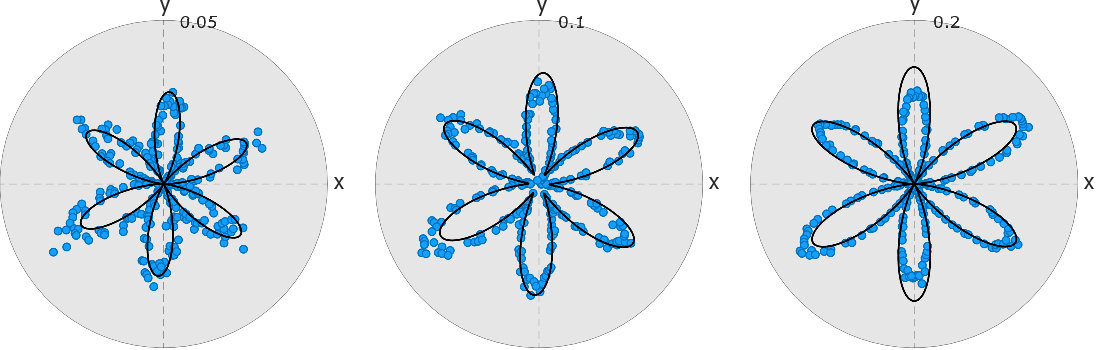}
	\caption{RA-SHG at 290 K for 1000 nm (left), 1005 nm (middle) and 1030 nm (right).  }
	\label{FigS1}
\end{figure}

At room temperature,  the magnitude of $\chi_{yyy}$ is negligible compared to $\chi_{xxx}$. As a result, the RA-SHG patterns at various wavelengths all possess nodes. Additionally, the intensity maxima are aligned to the crystal axes. In contrast, at lower temperatures, $\chi_{yyy}$ grows and according to Eq.~\ref{eq:S4}, the nodes can be lifted and the pattern can be rotated away from the crystal axes at various wavelengths depending on the $\chi_{xxx}/\chi_{yyy}$ ratio. In Fig.~\ref{FigS1}, we show three representative RA-SHG patterns at 290 K. 
The slight rotation away from the $y-$axis for the 1000~nm pattern indicates the presence of a residual $\chi_{yyy}$ contribution due to the ferro-rotational order. 
\newpage
\section{Temperature dependence of SHG at fixed wavelength}
In this section, we present the RA-SHG patterns at 1005 nm incident light, measured from 100 K to 290 K in 10 K increments.

\begin{figure}[hbt!]
	\includegraphics[width=0.95\columnwidth]{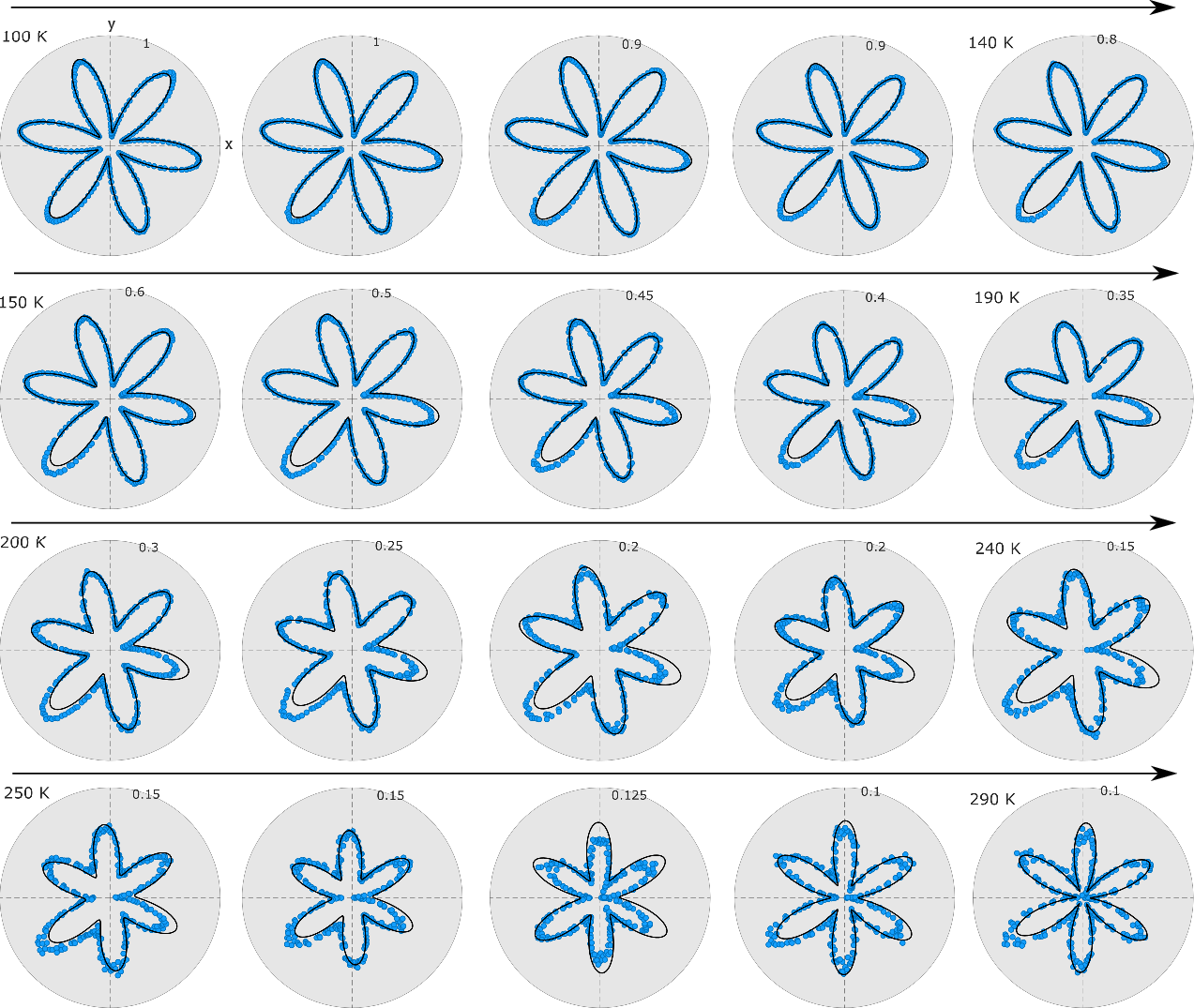}
	\caption{RA-SHG for 1005 nm incident wavelength  at various temperatures ranging from 100 K to 290 K. The top row displays RA-SHG from 100 K to 140 K, the second row from 150 K to 190 K, the third row from 200 K to 240 K and the bottom row from 250 K to 290 K (temperature increases from left to right for each row).   }
	\label{FigS2}
\end{figure}

\newpage
\section{Wavelength dependence of SHG at fixed temperature}
In this section, we show RA-SHG patterns at 100 K for various incident wavelengths ranging from 960~nm to 1030~nm. The fundamental wavelengths of 960-1030 nm correspond to the SHG energies of 2.4-2.6 eV, which resonate with several $d$-$d$ transitions in MnTiO$_3$. Across the crystal field transitions, the magnitude of $\chi_{xxx}$ and $\chi_{yyy}$ and the relative phase between them change drastically. Consequently, the patterns also change prominently with varying wavelength.
\begin{figure}[hbt!]
	\includegraphics[width=0.95\columnwidth]{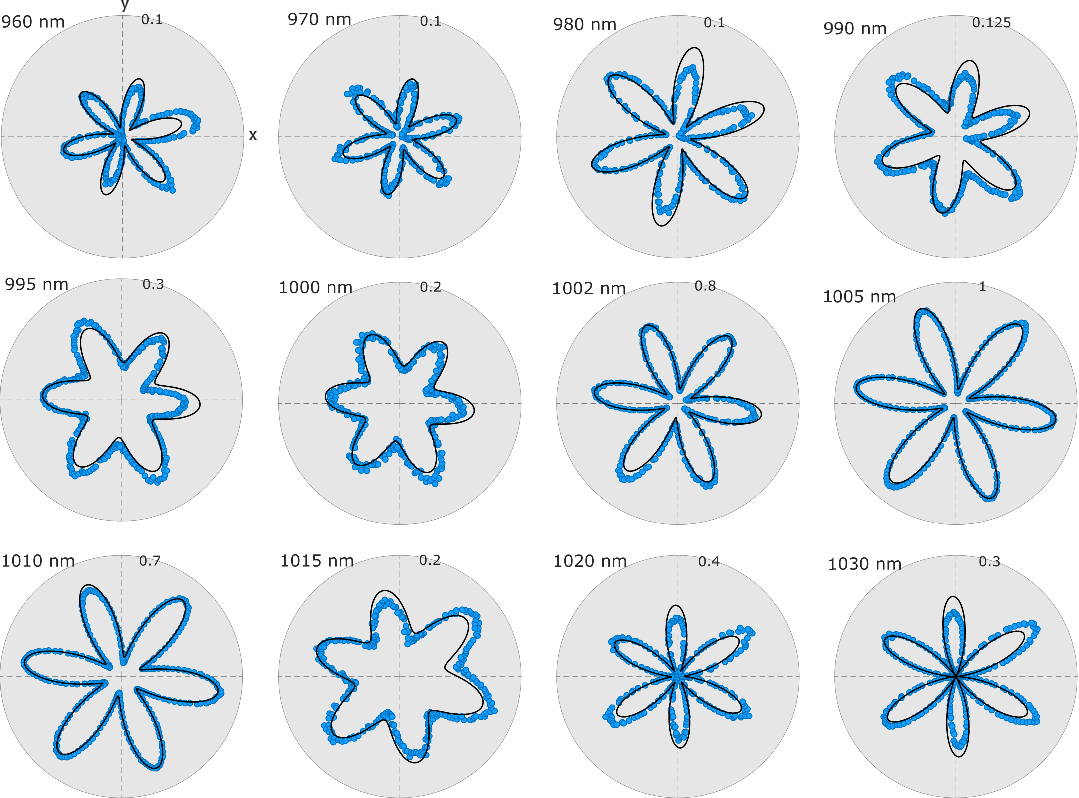}
	\caption{RA-SHG at 100 K for various wavelengths from 960 nm to 1030 nm.  }
	\label{FigS3}
\end{figure}
